\documentclass[a4paper]{jpconf}
\usepackage{graphicx}
\usepackage{amsmath}
\usepackage{epstopdf}
\usepackage[pdftex,colorlinks=true,linkcolor=black,citecolor=black,urlcolor=black,pdftitle={Matter neutrino oscillations without parametrization}, pdfauthor={Luis Jorge Flores Sandoval}, bookmarks=true]{hyperref}
\newcommand{\dm}[1]{\Delta m^2_{#1}}
\newcommand{\dM}[1]{\Delta M^2_{#1}}

\renewcommand\Re{\operatorname{Re}}
\renewcommand\Im{\operatorname{Im}}

\begin{document}
\title{Matter neutrino oscillations, an approximation in a
  parametrization-free framework}

\author{L. J. Flores~$^{1*}$ and O. G. Miranda~$^{1\dagger}$}

\address{$^1$~Departamento de F\'{\i}sica, Centro de
  Investigaci{\'o}n y de Estudios Avanzados del IPN\\ Apdo. Postal
  14-740 07000 Mexico, DF, Mexico}

\ead{$^*$jflores@fis.cinvestav.mx, $^\dagger$omr@fis.cinvestav.mx}

\begin{abstract}
Neutrino oscillations are one of the most studied and successful
phenomena since the establishment of the solar neutrino problem in 
late 1960's. 
In this work we discuss the exact expressions for the probability
$P_{\alpha\beta}$ in a constant density medium, in terms of the
standard vacuum parameters and the medium density. Besides of being
compact, these expressions are independent of any particular
parametrization, which could be helpful in the application of unitary
tests of the mixing matrix. In addition, we introduce a new
approximation on $P_{\alpha\beta}$ and compare it with the most
commonly used, discussing their main differences.
\end{abstract}

\section{Introduction}
Since the confirmation of the oscillation of neutrinos in 1998 by the
Super-Kamiokande collaboration~\cite{Fukuda:1998mi}, a considerable amount of
analysis has been made concerning the measurement of the parameters
involved in the oscillation~\cite{Forero:2014bxa}. In the standard
three families oscillation scheme, four parameters appear in the
mixing matrix: three mixing angles and a CP violation phase. The
former have been measured with great accuracy; however, the later had
not been properly constrained. Several
parameterizations for the leptonic mixing
matrix exist in the literature~\cite{Schechter:1980gr,Rodejohann:2011vc,Escrihuela:2015wra,deGouvea:2015euy,Agashe:2014kda}.
They are suited for different purposes; for example, to investigate
the nature of the neutrino (Dirac or Majorana), or to probe extra
neutrino states in oscillation
experiments~\cite{Escrihuela:2015wra,deGouvea:2015euy}. Most of the
analyses have been done with the Particle Data Group
parametrization~\cite{Agashe:2014kda}, but if we are looking to
perform a unitary test~\cite{Parke:2015goa}, it is desirable to use a
formulation as independent of the parametrization as
possible. Moreover, to obtain precise results, it is
necessary to take into account the matter effects on the
oscillation~\cite{Wolfenstein:1977ue}. The next generation of
experiments is looking to use this effect in future
analysis. Therefore, the analysis gets more complicated, requesting
the use of computational solutions or approximated expressions
\cite{Barger:1980tf,Zaglauer:1988gz,Cervera:2000kp,Freund:2001pn,Akhmedov:2004ny,Minakata:2015gra,Asano:2011nj}.\\ In
the first section of this work we give exact expressions for the
oscillation probability $P_{\alpha \beta}$ in a constant density
medium, which are formulated without a particular parametrization;
these will come in handy for future unitary tests. Additionally, in
the second section we introduce new approximated and simple
expressions, which can be defined to a desired degree of accuracy due
to their series expansion nature. Finally, we compare this
approximation with the most currently used and other two found in the
literature. It is important to notice that this work is based
on~\cite{Flores:2015mah}.

\section{The exact case}
The evolution equation for a neutrino $\nu_j$ with definite mass is given by 
\begin{equation}\label{1.1}
	i\frac{d}{dt}\nu_j=\frac{m_j^2}{2E}\nu_j.
\end{equation} 
\noindent These mass states are related to the flavor states
$\nu_\alpha$ through the leptonic mixing matrix, $U$, via $\nu_\alpha
= U_{\alpha j}\nu_j$, where a sum over $j$ is understood.  The mixing
matrix can be parametrized depending on the context of the study. From
(\ref{1.1}) one can calculate the known vacuum oscillation probability,
from $\nu_\alpha$ to $\nu_\beta$.
%
If extra heavy neutrinos exist, they can not participate in the
oscillation because the energy will not be sufficient, but the
unitarity of $U$ will be clearly affected. The $3\times 3$ mixing
matrix responsible of the oscillation is now just a block of the whole
$n\times n$ matrix $U$. This effect changes the usual $3\times 3$
expression into
\begin{eqnarray}\label{1.3}
	P_{\nu_\alpha \rightarrow \nu_\beta}=
\sum^3_{\ell ,j} U^*_{\alpha \ell}U_{\beta \ell}U_{\alpha j}U^*_{\beta j} - 
4\displaystyle\sum^{3}_{\ell >j} \Re\left[ U_{\alpha \ell}^{*}U_{\beta \ell}U_{\alpha j}U_{\beta j}^{*} \right] \sin^2\left( \frac{\Delta m^2_{\ell j}L}{4E}\right)& \nonumber \\ 
+2\displaystyle\sum^{3}_{\ell >j} \Im\left[ U_{\alpha \ell}^{*}U_{\beta \ell}U_{\alpha j}U_{\beta j}^{*} \right] \sin \left( \frac{\Delta m^2_{\ell j}L}{2E}\right)\,.&
\end{eqnarray}

\noindent Notice that now there is zero distance effect due to the
non-unitary mixing matrix. Additionally, if matter effects are to be
considered, we need to add the charge-current potential $V_{CC}$ to
the evolution equation in (\ref{1.1}) in the form of $A=2EV_{CC}$:
\begin{equation} \label{1.4}
	i\frac{d}{dt}\nu_j=\frac{1}{2E}\left( m_j^2 \nu_j + \displaystyle\sum_k AU_{e j}^* U_{ek}\nu_k\right).
\end{equation}

As usual, if we consider unitarity in $U$, the expression for the
probability in matter holds the same functional form as in vacuum;
with the replacement of $\Delta m^2_{\ell j}$ by the effective mass
difference, $\dM{\ell j }$, and the replacement of $U$ by the
effective mixing, V, that is related to the vacuum expression through
\begin{equation} \label{1.6}
	V=UW^T,
\end{equation}

\noindent where $W$ is the unitary matrix diagonalizing the
Hamiltonian in the right-hand side of equation (\ref{1.4}). One of the
purposes of this work is to express the matter probability as a
function entirely of the vacuum parameters and the
potential. Therefore our problem reduces to find the matrix $W$. This
can be done following the procedure of \cite{Zaglauer:1988gz} but with
the difference that we are leaving explicitly the elements $U_{\alpha
  i}$, in other words, without a specific parametrization. The Hamiltonian
\begin{equation} \label{1.7}
	H_M=\left(	\begin{array}{ccc}
 A|U_{e1}|^2 & A U_{e1}^* U_{e2} & A U_{e1}^* U_{e3} \\
A U_{e2}^* U_{e1} & \dm{21} + A|U_{e2}|^2 &  A U_{e2}^* U_{e3} \\
A U_{e3}^* U_{e1} & A U_{e3}^* U_{e2} & \dm{31} + A|U_{e3}|^2 \\
	\end{array}\right),
\end{equation}
\noindent has the characteristic polynomial form

\begin{equation}
\label{1.9}
	\lambda^3-\alpha \lambda^2 + \beta \lambda - \gamma=0,
\end{equation}

\noindent with
\begin{eqnarray} \label{1.11}
  \alpha &=& \dm{21}+\dm{31}+A (|U_{e1}|^2+ |U_{e2}|^2 + |U_{e3}|^2) 
  \nonumber \\
  \beta &=& \dm{31}\dm{21}+ A\dm{21}(|U_{e1}|^2 + |U_{e3}|^2)+
   A\dm{31}(|U_{e1}|^2+ |U_{e2}|^2) \nonumber \\
   \gamma &=& A\dm{21}\dm{31}|U_{e1}|^2 \\
   \eta &=& \cos\left[\frac{1}{3}\arccos\left( \frac{2\alpha^3-9\alpha\beta+27\gamma}{2\sqrt{(\alpha^2-3\beta)^3}}\right)\right] \nonumber .
\end{eqnarray}

\noindent Here we have left indicated on purpose the expressions
involving the condition of $U$ been unitary, to show how these
expressions can be easily changed in the case of a non-unitary study.
Now we can write the squared effective masses in terms of the previous
coefficients
\begin{eqnarray}\label{1.12}
	M_1^2 \equiv  \lambda_1 &=&  \frac{\alpha}{3}-\frac{1}{3}\sqrt{\alpha^2 -3\beta}\eta -\frac{\sqrt{3}}{3}\sqrt{\alpha^2 -3\beta}\sqrt{1-\eta^2}, \nonumber \\
	M_2^2 \equiv \lambda_2 &=&  \frac{\alpha}{3}-\frac{1}{3}\sqrt{\alpha^2 -3\beta}\eta +\frac{\sqrt{3}}{3}\sqrt{\alpha^2 -3\beta}\sqrt{1-\eta^2},  \\
	 M_3^2 \equiv \lambda_3 &=&  \frac{\alpha}{3}+\frac{2}{3}\sqrt{\alpha^2 -3\beta}\eta .  \nonumber 
\end{eqnarray}

The columns that constitute the diagonalizing matrix $W$ correspond to
the eigenvectors of $H_M$. A suitable choice of them gives us
\begin{equation}\label{1.16}
(W^T)_{k j}= \frac{\Lambda_k}{C_k}\delta_{k j}
+(1-\delta_{k j})A \frac{U_{ek}U_{ej}^*\left(M_k^2-\sum_i[\dm{i1}\epsilon^2_{ijk}] \right)}{C_k}\,,  
\end{equation}

\noindent where we have defined 
\begin{equation}\label{1.14}
	\Lambda_j= M_j^4- \displaystyle\sum_{i\neq j}\left[M_j^2\left( \dm{i1}+A|U_{ei}|^2\right)-A \dm{i1} |U_{ek}|^2-\frac{1}{2}\dm{i1} \dm{k1}\right], \quad \mbox{for} \quad k\neq i ,
\end{equation}
\noindent and a normalization constant, $C_j$, as
\begin{equation}\label{1.15}
C_j=\sqrt{\Lambda_j^2+A^2|U_{ej}|^2\displaystyle\sum_{i\neq j}|U_{ei}|^2(M_j^2-\dm{k1})^2},\quad \mbox{for}\quad k\neq i .
\end{equation}

With the help of the relation~(\ref{1.6}) we are able to write down a
compact expression, relating the entries of the effective matter
mixing matrix $V$:
\begin{equation}\label{1.17}
			V_{\beta j}= \frac{\Lambda_j}{C_j}U_{\beta j}+A\displaystyle\sum_{k\neq j}\frac{U_{\beta k}U_{ei}U_{ej}^*\left(M_k^2-\sum_i[\dm{i1}\epsilon^2_{ijk}] \right)}{C_k}.
		\end{equation}	
 
\noindent It is easy to note that this expression reduces to the
vacuum case when $A=0$. This relation is useful in the description of
neutrino matter oscillation probabilities as a function of the
vacuum parameters, without a particular choice of parametrization
for the leptonic mixing matrix. The unitarity of $U$ is not assumed. 
That makes it useful in case a unitary test would be implemented.

\section{Approximated expressions}

Now that we have presented the exact formula for the neutrino
oscillation probability, free of any parametrization, we will start
with the process of deducing an approximated formula. Although there
exist several approximations in the literature, our expression will
stick to the purpose of a formulation without parametrization. To
accomplish this, we are looking to approximate the effective masses
appearing in (\ref{1.12}).\\
First, we notice that $\dm{21}=0$ implies $\gamma =0$, and the
polynomial in (\ref{1.9}) reduces to a quadratic form. If this is the
case, we have:
\begin{equation}
\eta  \label{1.18}
= \cos\left[\frac{1}{3}\arccos\left( \frac{2\alpha^3-9\alpha\beta}{2\sqrt{(\alpha^2-3\beta)^3}}\right)\right] \,,
\end{equation}
 \noindent which reduces to
\begin{equation}
\eta \label{1.19}
= \cos\theta 
= \frac{-\frac{1}{2}\alpha}{\sqrt{\alpha^2-3\beta}} .
\end{equation}
\noindent Now we go to the real case: it is well known that $\dm{21}\ll \dm{31}$, therefore, $\gamma$ is different from zero and $\gamma \ll \alpha \beta$. We suggest a correction $\varepsilon$ that needs to fulfill
\begin{equation}
\cos\theta \label{1.20}
= \frac{-\frac{1}{2}\alpha + \varepsilon}{\sqrt{\alpha^2-3\beta}}
\end{equation}
and 
\begin{equation} \label{1.21}
\cos3\theta = 4\cos^3\theta - 3 \cos\theta \simeq  
\frac{2\alpha^3-9\alpha\beta+27\gamma}{2\sqrt{(\alpha^2-3\beta)^3}} . 
\end{equation}
\noindent Working up to first-order terms in $\gamma / \beta$ is easy to find that
$\varepsilon = \frac{3 \gamma}{2\beta}$
fulfills both conditions (\ref{1.20}) and (\ref{1.21}). The eigenvalues in (\ref{1.12}) are then approximated to
\begin{eqnarray}\label{1.23}
	 M_1^2 \equiv \lambda_3 &\simeq&  \frac{2}{3}\varepsilon.  \nonumber \\
M_2^2 \equiv  \lambda_1 &\simeq&  
         \frac{1}{2}\left(\alpha-\frac{2}{3}\varepsilon\right) 
       -\frac{1}{2}\sqrt{\left(\alpha+\frac{2}{3}\varepsilon\right)^2 
         -4\left[\beta+\left(\frac{2}{3}\varepsilon\right)^2\right]},  \\
M_3^2 \equiv \lambda_2 &\simeq&  
         \frac{1}{2}\left(\alpha-\frac{2}{3}\varepsilon\right) 
       +\frac{1}{2}\sqrt{\left(\alpha+\frac{2}{3}\varepsilon\right)^2 
       -4\left[\beta+\left(\frac{2}{3}\varepsilon\right)^2\right]}, \nonumber
\end{eqnarray}
\noindent If we substitute these back in the polynomial in (\ref{1.9}), it takes the form:
\begin{equation}
\label{1.24}
	\lambda^3-\alpha \lambda^2 + \beta \lambda 
        - \left(\frac{2}{3}\beta\varepsilon - \frac{4}{9} \alpha\varepsilon^2+
          \frac{8}{27}\varepsilon^3\right)=0 .
\end{equation} 
We want to go beyond this and find an expression for $\varepsilon$ at second order in $\gamma / \beta$. Therefore, we propose 
\begin{equation}
\label{1.26}
\varepsilon = \frac{3\gamma}{2\beta} + a_2\left(\frac{\gamma}{\beta}\right)^2\,.
\end{equation}
\noindent Substituting this in~(\ref{1.24}), and demanding that
returns to the form in (\ref{1.9}) up to second-order terms, throws the
result $a_2= \frac{3}{2}\frac{\alpha}{\beta}$. Hence, we have
\begin{equation}
\label{1.27}
\varepsilon = \frac{3 \gamma}{2\beta} + \frac{3\alpha}{2\beta}\frac{\gamma^2}{\beta^2} .
\end{equation}
 We can go further with this method and find the coefficients $a_k$ in
 a recursive way. This states that we can write the
 correction $\varepsilon$ as an infinite series in powers of $\gamma /
 \beta$ and, in principle, obtain the exact case again. The correction
 then takes the form
\begin{equation}
\label{1.28}
\varepsilon = \sum_{k=1}^\infty a_k(\alpha,\beta) \left[\frac{\gamma}{\beta}\right]^k 
\end{equation}
with
\begin{equation}
a_1 = \frac{3}{2}, \quad  \phantom{...}  
a_k(\alpha,\beta) = \frac{3}{2\beta}
             \left(\frac{4\alpha}{9} \sum_{\substack{i,j\\i+j=k}}a_ia_j 
              -\frac{8}{27} \sum_{\substack{i,j,l\\i+j+l=k}}a_ia_j a_l \right) ; 
  \,\,\, k\geq 2.
\end{equation}

\begin{figure}[h]
\begin{minipage}{18pc}
\includegraphics[width=18pc]{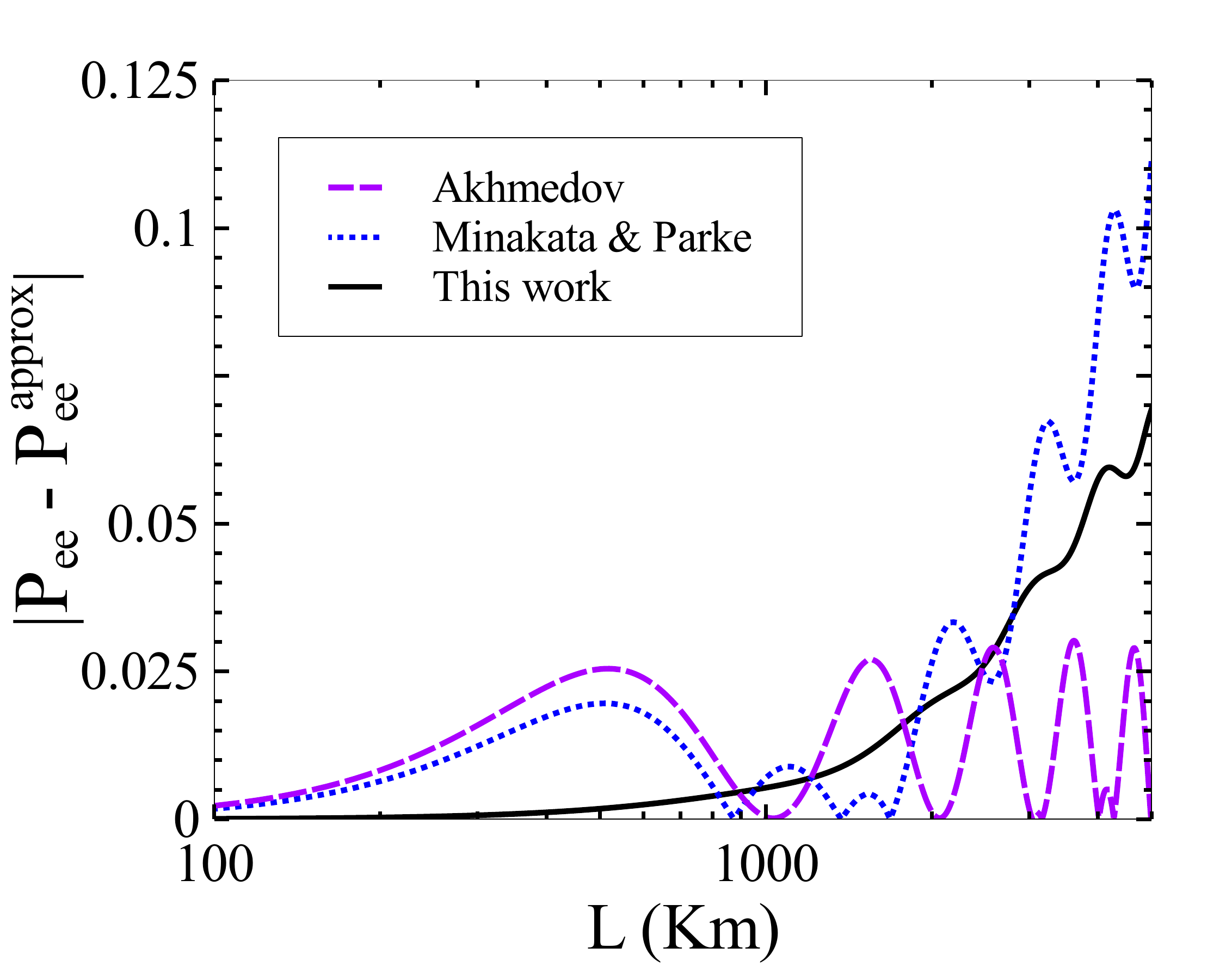}
\end{minipage}\hspace{2pc}%
\begin{minipage}{18pc}
\includegraphics[width=18pc]{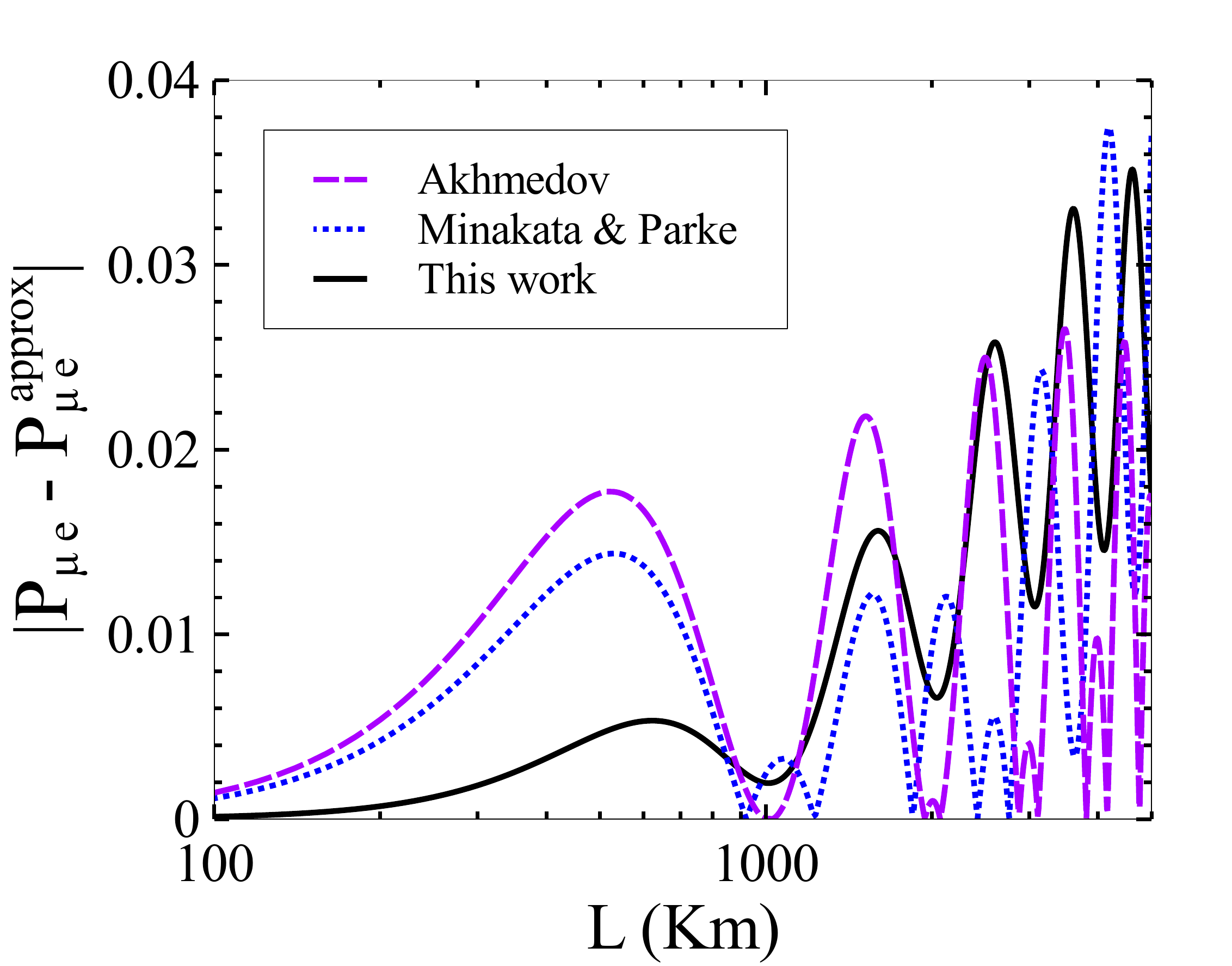}
\end{minipage} 
\caption{Comparison between the exact neutrino oscillation probability in matter and some approximations. The absolute difference for the electron neutrino survival probability $P_{ee}$ (left panel) and the the muon to electron probability $P_{\mu e}$ (right panel) are plotted against the distance $L$ at an energy of $E_\nu = 1$~GeV. The central values of the mixing matrix parameters and the square mass differences are taken from \cite{Forero:2014bxa} and an electron density of $N_e=5.92\times 10^9$~eV$^3$ has been assumed. The dashed magenta curve accounts for the approximation by Akhmedov \cite{Akhmedov:2004ny} as well as the dotted blue curve stands for the expression from Minakata \& Parke \cite{Minakata:2015gra}. The solid curve represents the difference with our approximation at first order.}
\label{AllDifferAbs}
\end{figure}

\begin{figure}[h]
\begin{minipage}{18pc}
\includegraphics[width=18pc,height=0.21\textheight]{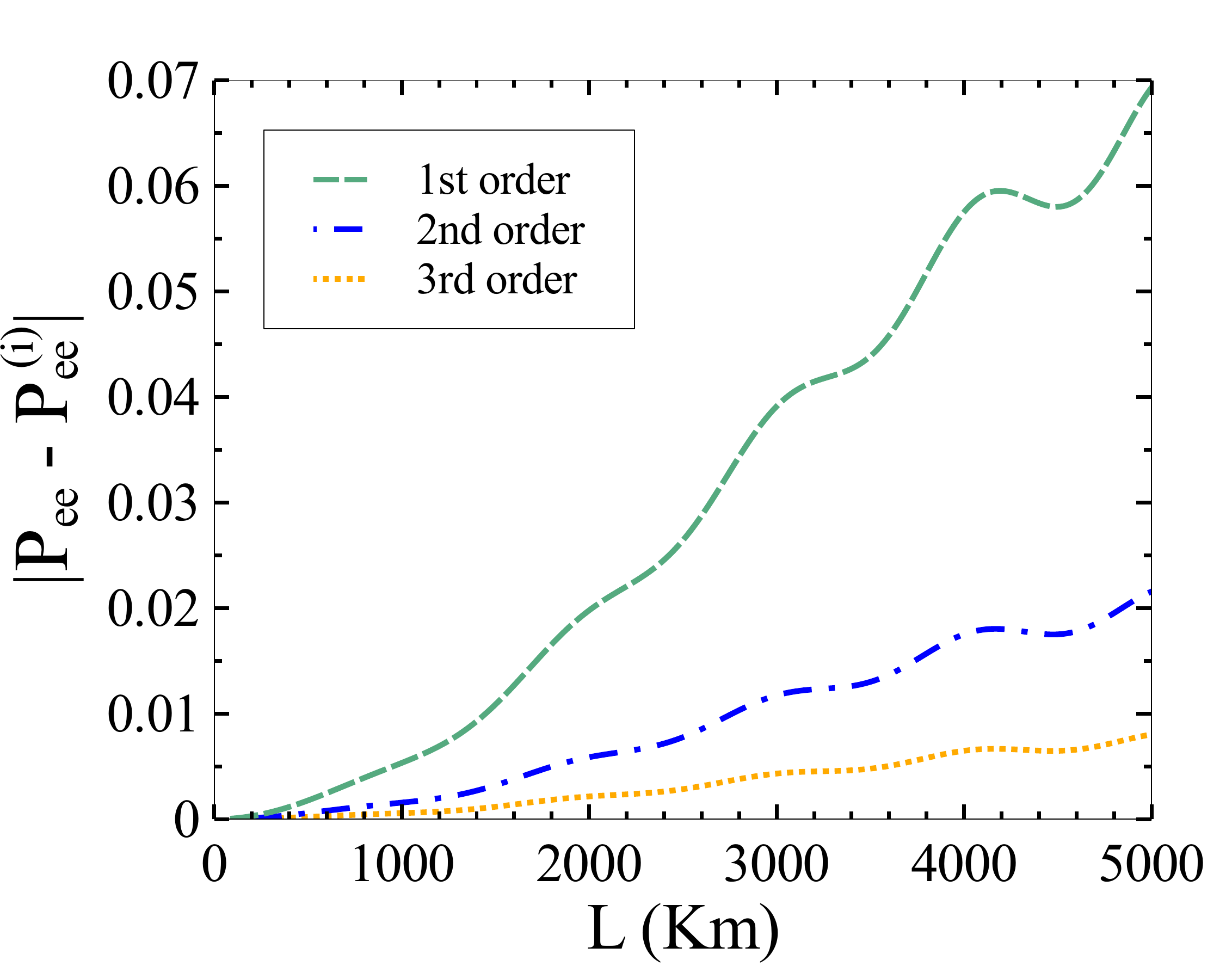}
\end{minipage}\hspace{2pc}%
\begin{minipage}{18pc}
\includegraphics[width=18pc,height=0.21\textheight]{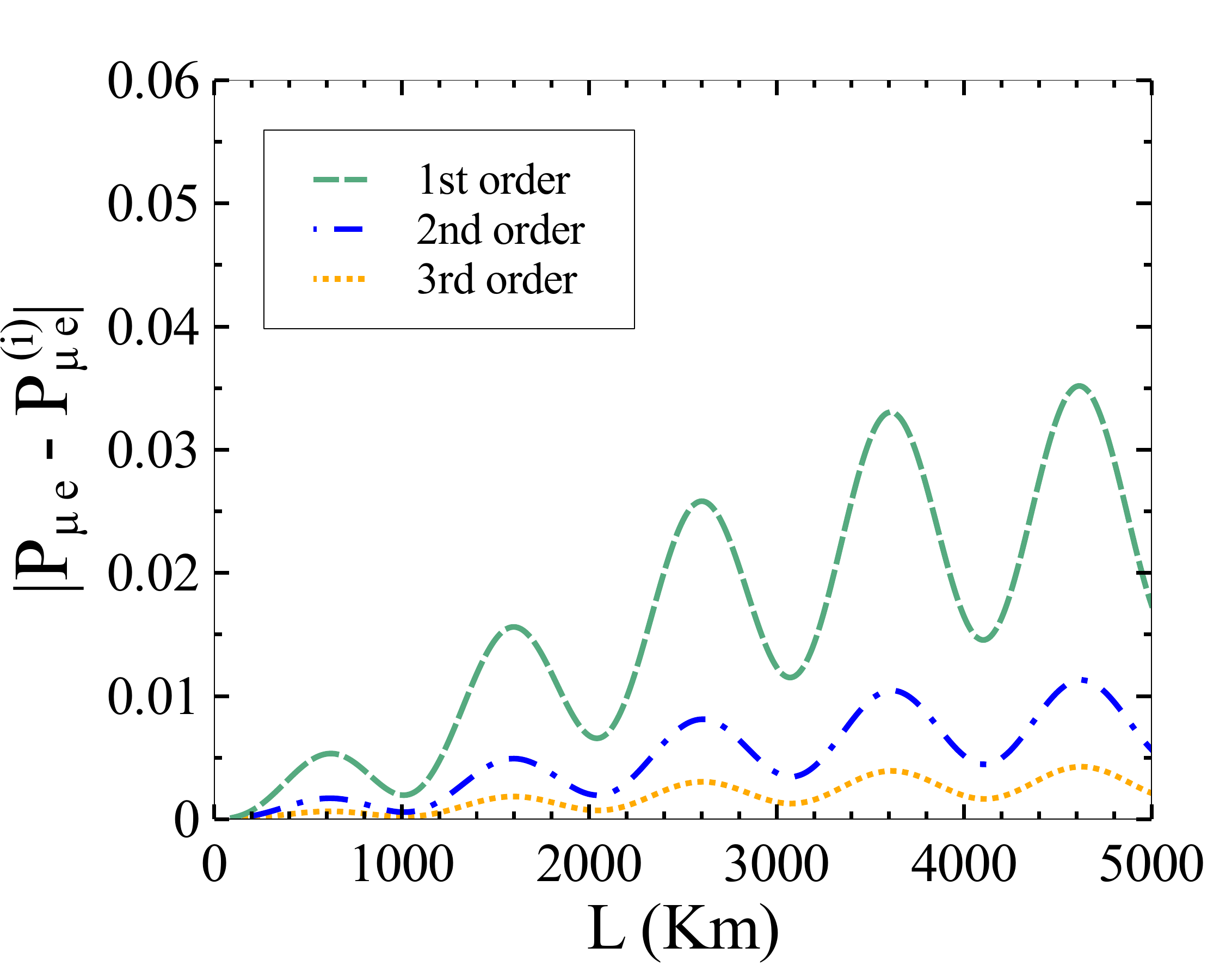}
\end{minipage} 
\caption{Comparison between the exact formula and our approximation for the neutrino oscillation probability. The absolute differences for $P_{ee}$ (left panel) and $P_{\mu e}$ (right panel) are plotted against the distance $L$, at an energy of $E_\nu=1$~GeV. The dashed, dash-dotted and dotted line correspond to the first, second and third order approximation, respectively.}
\label{DifferOrders}
\end{figure}

Now that we have defined the approximation, we would like to test its
accuracy by comparing it against the exact formula and also to other
known approximations: by Akhmedov \cite{Akhmedov:2004ny} and Minakata
\& Parke \cite{Minakata:2015gra}. To do so, we first assume unitarity
in the mixing matrix and adopt the standard parametrization. The
central values for the mixing parameters are the reported ones in
\cite{Forero:2014bxa} ($\sin^2\theta_{12}=0.320$,
$\sin^2\theta_{23}=0.613$, $\sin^2\theta_{13}=0.0246$) along with the
squared mass differences ($\dm{21}=7.62\times 10^{-5}$~eV$^2$,
$\dm{31}=2.55\times 10^{-3}$~eV$^2$). We have taken $\delta = 3\pi /
2$ as the value of the CP phase. We computed the oscillation
probabilities $P_{ee}$ and $P_{\mu e}$ as function of the baseline at
an energy of $1$~GeV. In Fig.~(\ref{AllDifferAbs}) we compare our
result with two other approximations from \cite{Akhmedov:2004ny,
  Minakata:2015gra}; we can see that our result is competitive with
the others, in particular for a small distance (short-baseline
experiments). The precision between different orders of our
approximation is shown in Fig.~(\ref{DifferOrders}). Let us notice
that, at first order approximation, our result works well,
especially for a short baseline and a great development is achieved for
higher orders.

\section{Conclusions}
	In this work, we studied the scenario of the oscillation of
        three neutrino families propagating through a constant
        potential caused by matter. First we reconsidered the exact
        expressions for the oscillation probability, but without any
        parametrization. Additionally, we found an approximation for
        these expressions, using the fact that $\dm{21}\ll \dm{31}$
        and keeping thus the parametrization-free scheme. We found
        that this result could be expanded to any order in a series
        expansion, making simple the choice of required precision. Our
        result is competitive with the ones found in the literature,
        and its value goes closer to the exact case for short
        baselines.

\section*{Acknowledgments}
This work has been supported by the CONACyT Grant No. 166639 (Mexico).

\end{document}